\documentclass[useAMS,usenatbib]{mn2e}
\usepackage{times,graphicx}
\input{epsf}

\defcitealias{vaz08}{V08}

\title[From round to multipolar]
{NGC\,6309, a planetary nebula that shifted from round to multipolar\thanks{
Based on observations made with the Nordic Optical Telescope, operated by the Nordic Optical Telescope Scientific Association at the Observatorio del Roque de los Muchachos, La Palma, Spain, of the Instituto de Astrof'\'{\i}sica de Canarias}\thanks{Based upon observations acquired at the Observatorio Astron\'omico Nacional in the Sierra San Pedro M\'artir (OAN-SPM), Baja California, Mexico.}}

\author[Rubio, et al.]
{G. Rubio$^{1}$\thanks{E-mail:gerardo@astro.iam.udg.mx},
R. V\'{a}zquez$^{2}$, G. Ramos-Larios$^{3}$, M. A. Guerrero$^{4}$,  
L. Olgu\'{\i}n$^{5}$ \newauthor P. F. Guill\'{e}n$^{2}$, H. Mata$^{1}$\\
$^{1}$CUCEI, Universidad de Guadalajara, 
Blvd. Marcelino Garc'\'{\i}a Barrag\'an 1421, Guadalajara, Jalisco, Mexico\\
$^{2}$Instituto de Astronom\'{\i}a, Universidad Nacional Aut\'onoma de
M\'exico, Apdo. Postal 877, 22800 Ensenada, B. C., Mexico\\
$^{3}$Instituto de Astronom\'{\i}a y Meteorolog\'{\i}a, CUCEI, Universidad de Guadalajara, Av. Vallarta No. 2602, Col. Arcos Vallarta, 44130 Guadalajara, Jalisco, Mexico\\
$^{4}$Instituto de Astrof\'{\i}sica de Andaluc\'{\i}a, IAA-CSIC, C/Glorieta de la Astronom'\'{\i}a s/n, 18008 Granada, Spain\\
$^{5}$Depto. de Investigaci\'on en F\'{\i}sica, Universidad de Sonora, Blvd. Rosales Esq. L. D. Colosio, Edif. 3H, 83190 Hermosillo, Sonora, Mexico\\}

\begin{document}

\date{Received 2014 September 01}

\pagerange{\pageref{firstpage}--\pageref{lastpage}} \pubyear{}

\maketitle

\label{firstpage}
\begin{abstract} 

We present new narrow-band H$\alpha$, [N~{\sc ii}], and [O~{\sc iii}]  
high-resolution images of the quadrupolar planetary nebula (PN) NGC\,6309 
that show in great detail its bipolar lobes and reveal new morphological 
features.  
New high- and low-dispersion long-slit spectra have been obtained to 
help in the investigation of the new nebular components.  
The images and spectra unveil two diffuse blobs, one of them located 
$\simeq55\arcsec$ from the central star along the NE direction (PA=+71\degr) 
and the other at $\simeq78\arcsec$ in the SW direction (PA=--151\degr). 
Therefore, these structures do not share the symmetry axes of the 
inner bipolar outflows.  
Their radial velocities relative to the system are quite low: +3 and --4 
km\,s$^{-1}$, respectively.
Spectroscopic data confirm a high [O~{\sc iii}] to H$\beta$ ratio, 
indicating that the blobs are being excited by the UV flux from 
the central star. 
Our images convincingly show a spherical halo 60\arcsec\ in diameter 
encircling the quadrupolar nebula.  
The expansion velocity of this shell is low, $\leq$6 km~s$^{-1}$.  
To study the formation history of NGC\,6309, we have used our new images 
and spectra, as well as available echelle spectra of the innermost regions, 
to estimate the kinematical age of each structural component: 
the software {\sc shape} has been used to construct a morpho-kinematic 
model for the ring and the bipolar flows that implies an age of $\sim$4,000 
yrs, the expansion of the halo sets a lower limit for its age $\geq$46,000 
yrs, and the very low expansion of the blobs suggests they are part of a 
large structure corresponding to a mass ejection that took place $\sim$150,000 
yrs ago.  
In NGC\,6309 we have direct evidence of a change in the geometry of 
mass-loss, from spherical in the halo to axially-symmetric in the 
two pairs of bipolar lobes.  
 
\end{abstract}


\section{Introduction}

\begin{figure*}
\begin{center}
\includegraphics[width=0.8\textwidth,angle=0]{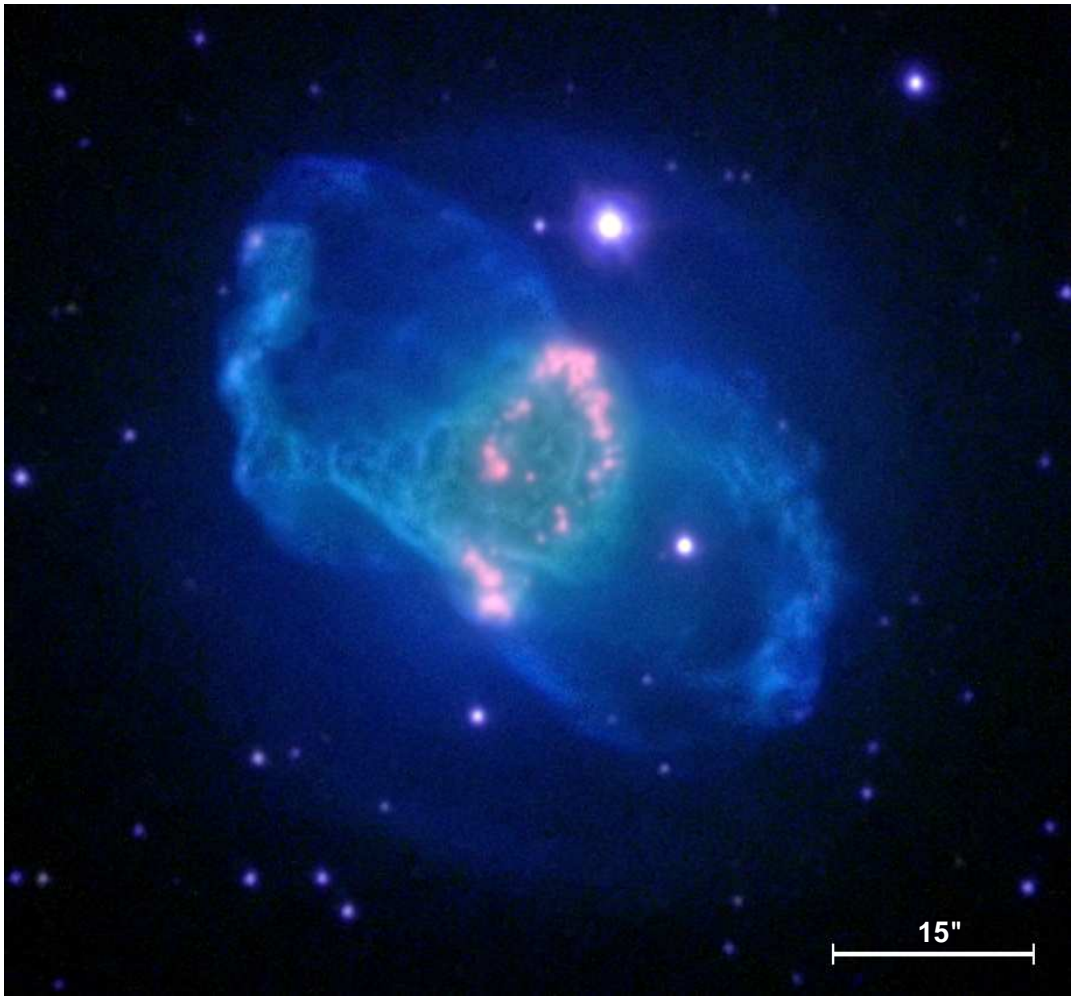}
\caption{Colour-composite NOT picture of NGC\,6309 in the [N~{\sc ii}] (red), 
H$\alpha$ (green), and [O~{\sc iii}] (blue) emission lines.  
This new picture reveals into great detail the fragmented emission of 
[N~{\sc ii}] from the central ring, the bipolar lobes, and the outer 
round shell (The reddish colour of the central star is a processing artefact).  
North is up, east to the left.
\label{NOTzoom}
}
\end {center}
\end{figure*}

Planetary nebulae (PNe) are the progeny of low- and intermediate-mass 
stars ($0.8-1.0 M_\odot \leq M_i \leq 8-10 M_\odot$).  
The Interacting Stellar Winds (ISW) model has been the canonical scenario 
to explain their formation and evolution \citep{kwo78}.  
In the ISW model, the formation of bipolar or axi-symmetric PNe requires 
the presence of a density gradient between the equatorial and polar 
directions \citep{bal87}.  
The discovery of highly elongated PNe \citep[e.g., KjPn\,8,][]{lop95}, 
high velocity outflows \citep[e.g., NGC\,2392,][]{gie85}, and PNe with 
multiple symmetry axes \citep[e.g., the quadrupolar PNe,][]{man96}, 
has caused many problems for the ISW model \citep{bal02}.  

In this sense, there have been proposals of alternative or additional 
mechanisms that contribute to the formation and evolution of PNe, 
particularly for the ejection of precessing highly-collimated outflows.  
These models include the interaction of a binary system \citep{sok94} 
that may give rise to accretion of material onto an equatorial disc, 
or the action of strong magnetic fields \citep{gar99}.  
Whatever the actual mechanism is, it is generally accepted that the 
onset of asymmetry occurs at the end of the Asymptotic Giant Branch 
(AGB) or in the short transition from the AGB to the post-AGB phase.  
Collimated outflows seem to play a significant role in this process 
\citep{sah98}.

Quadrupolar and multipolar PNe and those exhibiting collimated outflows 
at different directions are key, as precession can be linked to the 
final evolution of its progenitor star and its interaction with a binary 
companion or the occurrence of magnetic fields \citep{cor14,toc14}.  
One singular object is NGC\,6309 (a.k.a.\ the Box Nebula), a recently 
recognized quadrupolar PN consisting of a central ring and two pairs 
of bipolar lobes aligned along different directions \citep[][hereafter 
V08]{vaz08}.  
Interestingly, there were also hints for the presence of an outer structure 
around these bipolar lobes, defined as a halo by V08 in the basis of their 
narrow-band images.  
Additional evidence for this outer shell is found in the spectroscopic 
echelle observations of \citet{chu89} who reported the presence of a 
slowly expanding outer shell.  

In this paper we revisit NGC\,6309 to confirm the presence of the outer 
structures hinted by V08.  
Deep, high-resolution narrow-band images have allowed us to find a 
round halo that encircles the inner quadrupolar lobes.  
Much fainter nebular emission is found at large distances from the 
central star.  
Low- and high-dispersion optical spectra have been used to investigate 
the nature of these morphological components.  
The paper is organized as follows:  
Sec.\,2 describes the new observations, the results are presented 
in Sec.\,3, and the nature of the new features and the formation 
history of the nebula are discussed in Sec.\,4.  
A final summary is presented in Sec.\,5.  

\begin{figure*}
\begin{center}
\includegraphics[width=1\textwidth]{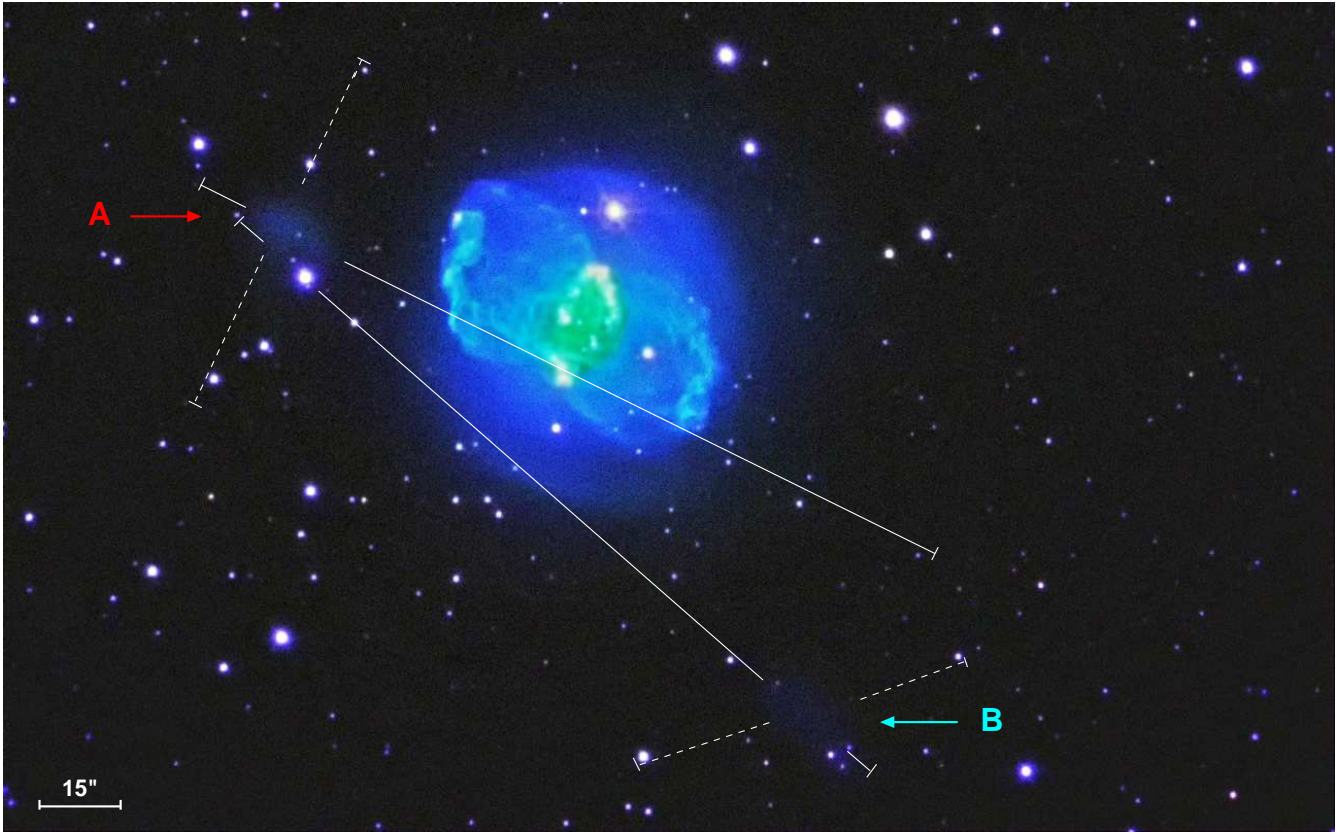} 
\caption{
Same as Figure~1, but for a larger field of view.  
The outer blobs ``A'' and ``B'' are labeled.  
The dotted lines mark the position of the slits used for low-dispersion 
spectroscopy, whereas the solid lines mark that of the high-dispersion 
spectra, corresponding at 160$\arcsec$ length.
}
\label{NOTlarge}
\end{center}
\end{figure*}

\section{Observations}
\label{observations}

\subsection{High-resolution imagery}
\label{imagenes}

New narrow-band optical images of NGC\,6309 were obtained on 2009 July 
21 using the Andalucia Faint Object Spectrograph and Camera (ALFOSC) at 
the 2.5-m Nordic Optical Telescope (NOT) of the Roque de los Muchachos 
Observatory (ORM, La Palma, Spain).  
The images were acquired through narrow-band filters that isolate the 
H$\alpha$ (\mbox{$\lambda_c = 6567$\AA}, \mbox{FWHM $= 8$\AA )},
[N~{\sc ii}] (\mbox{$\lambda_c = 6588$\AA}, \mbox{FWHM $= 9$\AA)}, and
[O~{\sc iii}] (\mbox{$\lambda_c = 5007$\AA}, \mbox{FWHM $= 30$\AA)} 
emission lines.  
The detector was a 2048$\times$2048 EEV CCD with a pixel size of 13.5~$\mu$m, 
which implies a plate scale of 0\farcs184~pix$^{-1}$ and a field of view of 
6\farcm6$\times$6\farcm6.  
Exposure times were 900\,s, 900\,s, and 600\,s for the H$\alpha$, 
[N\,{\sc ii}], and [O\,{\sc iii}] images, respectively (Table~1). 
The images were processed using standard {\sc iraf}\footnote{IRAF is distributed by the National Optical Astronomy Observatory, which is operated by the Association of Universities for Research in Astronomy (AURA) under cooperative agreement with the National Science Foundation.} routines.  
Composite colour-pictures of these images are shown in Figure~\ref{NOTzoom} 
and Figure~\ref{NOTlarge}.  
The mean spatial resolution, as determined from the FWHM of stars in 
the FoV, was 0\farcs7.  

A deep [O~{\sc iii}] image was obtained on 2014 July 18 also using ALFOSC 
at the NOT.  
This time, the EEV CCD pixels were binned by three into pixels 
corresponding to a plate scale of 0\farcs552.  
Three 200\,s and three 450\,s exposures were obtained for a total 
integration time of 1,950\,s  (Table~1) in the search for faint 
outer structures.  
The image was similarly processed using standard {\sc iraf} routines and 
the spatial resolution implied by field stars was alike, $\sim$0\farcs8.

\begin{table}\centering
\setlength{\columnwidth}{0.2\columnwidth}
\setlength{\tabcolsep}{1.30\tabcolsep}
\caption{NOT ALFOSC optical observations of NGC\,6309}
\begin{tabular}{lrrccr}
\hline
\hline 
\multicolumn{1}{c}{Filter} & 
\multicolumn{1}{c}{$\lambda$  $_{c}$} & 
\multicolumn{1}{c}{$\Delta\lambda$} & 
\multicolumn{1}{c}{Exp.\ Time} &
\multicolumn{1}{c}{Frames} &
\multicolumn{1}{c}{Date}\\
\multicolumn{1}{c}{} & 
\multicolumn{1}{c}{(\AA)} & 
\multicolumn{1}{c}{(\AA)} & 
\multicolumn{1}{c}{(s)} &
\multicolumn{1}{c}{} &
\multicolumn{1}{c}{}\\
\hline  
\phantom{x}[O\,{\sc iii}]  &  5007 &  30 & 300 & 2 & July 21th, 2009 ~~\\
\phantom{x}H$\alpha$       &  6567 &   8 & 450 & 2 & July 21th, 2009 ~~\\
\phantom{x}[N\,{\sc ii}]   &  6588 &   9 & 450 & 2 & July 21th, 2009 ~~\\
\phantom{x}[O\,{\sc iii}]  &  5007 &  30 & 200 & 3 & July 18th, 2014 ~~\\
\phantom{x}[O\,{\sc iii}]  &  5007 &  30 & 450 & 3 & July 18th, 2014 ~~\\
    \hline
  \end{tabular}
\vspace{0.5cm}
\label{logobs}
\end{table} 

\subsection{Long-slit low-dispersion spectroscopy}

Low-resolution, long-slit spectra were obtained with the Boller \& Chivens spectrograph 
mounted on the 2.1m telescope at the Observatorio Astron\'omico Nacional in the Sierra San Pedro M\'artir 
(OAN-SPM, Baja California, Mexico) on 2014 May 20 and 21. 
An E2V CCD (13.5 $\mu$m~pix$^{-1}$) with 2048$\times$2048
pixel array was used as detector. 
The 400 lines\,mm$^{-1}$ grating was used along with a 2\farcs6 and 3\farcs9 slit 
width, yielding a spectral resolution of 4.2\,{\AA} and 5.2\,{\AA} (FWHM) respectively. 
We have obtained three spectra on each position with an exposure time of 
1800 s for each spectra, and combined them to get the median spectrum.
The two slit position are indicated in Figure~\ref{NOTlarge} at PA=+155\degr and PA=+107\degr.
The first slit covers the NE blob (hereafter blob A) and the second
the SW blob (hereafter blob B).
In both cases the brighter part of the emission region in 
[O\,{\sc iii}]$\lambda$5007 was included. 
The median spectrum was then subtracted from individual spectrum to detect and eliminate  
cosmic rays manually. Cleaned \textit{spectra} were then added in order to get the final spectrum.
Spectra reduction was carried out following standard procedures in 
XVISTA\footnote{XVISTA was originally developed as Lick Observatory Vista. It is currently 
maintained by Jon Holtzman at New Mexico State University and is available at
http://ganymede.nmsu.edu/holtz/xvista.}.

\subsection{Long-slit echelle spectroscopy}

Two high-resolution, long-slit spectra were obtained on 2013 May 19 with the Manchester Echelle 
Spectrograph (MES; \citealt{mea03}) in the 2.1-m telescope at OAN-SPM. 
An E2V 13.5-$\mu$m\,pix$^{-1}$ 
CCD with 2048$\times$2048 pixels was used as detector in 4$\times$4 binning mode 
resulting in a spatial scale of 0\farcs702 \,pix$^{-1}$ and a spectral scale of 0.087 {\AA}\,pix$^{-1}$. 
The spectra were centred at the [O\,{\sc iii}]$\lambda5007$ 
emission line using a filter ($\Delta\lambda$=60\AA) to isolate the 114$^{\rm th}$ order. Exposure time was 1800\,s for each spectrum. 
The two slit position are indicated in Figure~\ref{NOTlarge} at PA=64\degr (NE, solid line) and 
PA=+50\degr (SW, solid line).

Data were calibrated using {\sc iraf} standard procedures for long-slit 
spectroscopy. 
The slit width
was 150\,$\mu$m (1.9$\arcsec$). 
The resulting spectral resolution (FWHM) is $\simeq$12 km~s$^{-1}$ 
(accuracy $\pm$1 km~s$^{-1}$), as measured from the lines of the 
ThAr calibration lamp.  
Seeing was $\sim$2\arcsec\ during the observations.

\begin{figure}
\begin{center}
\includegraphics[width=0.45\linewidth]{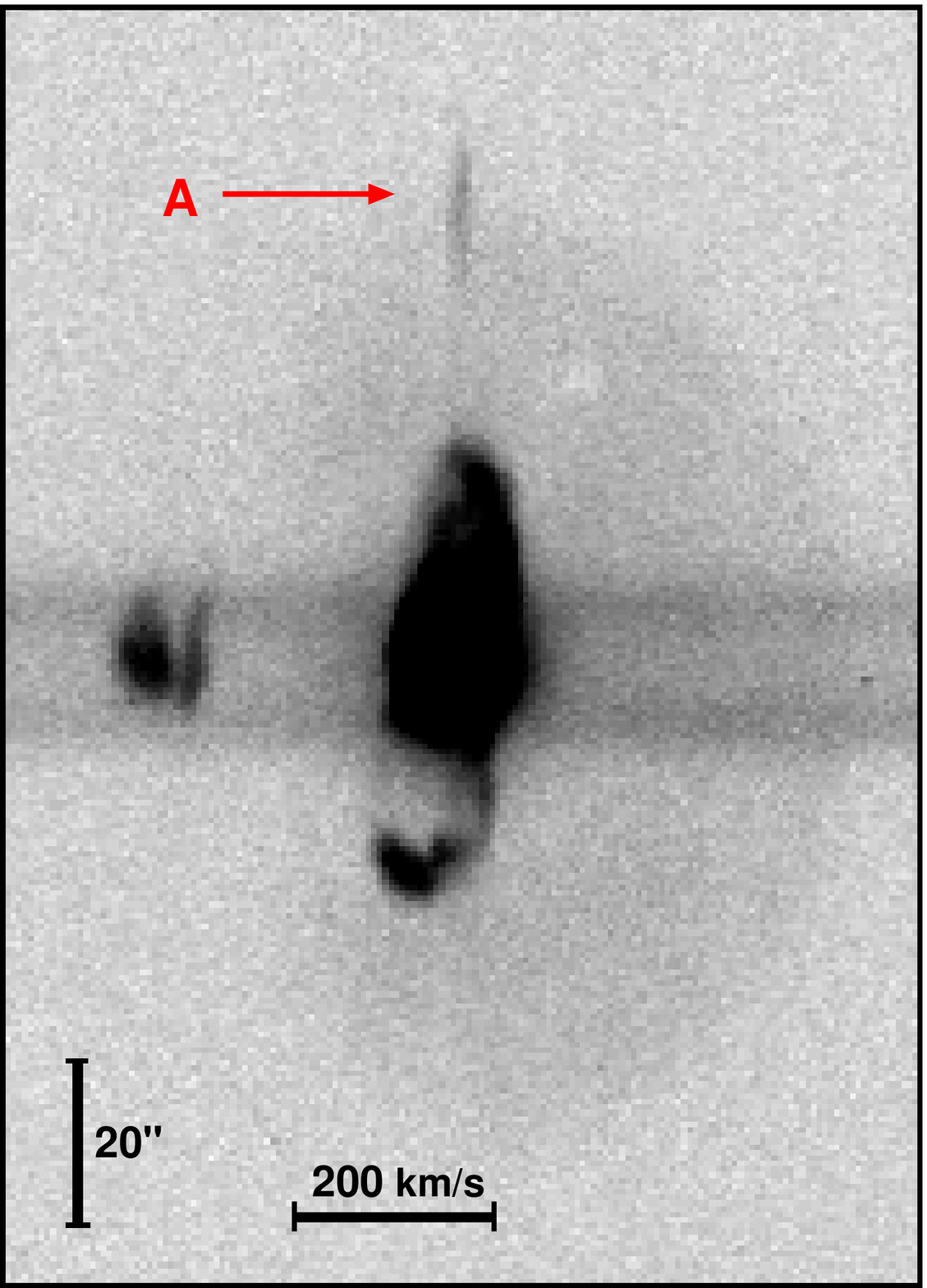}
\includegraphics[width=0.452\linewidth]{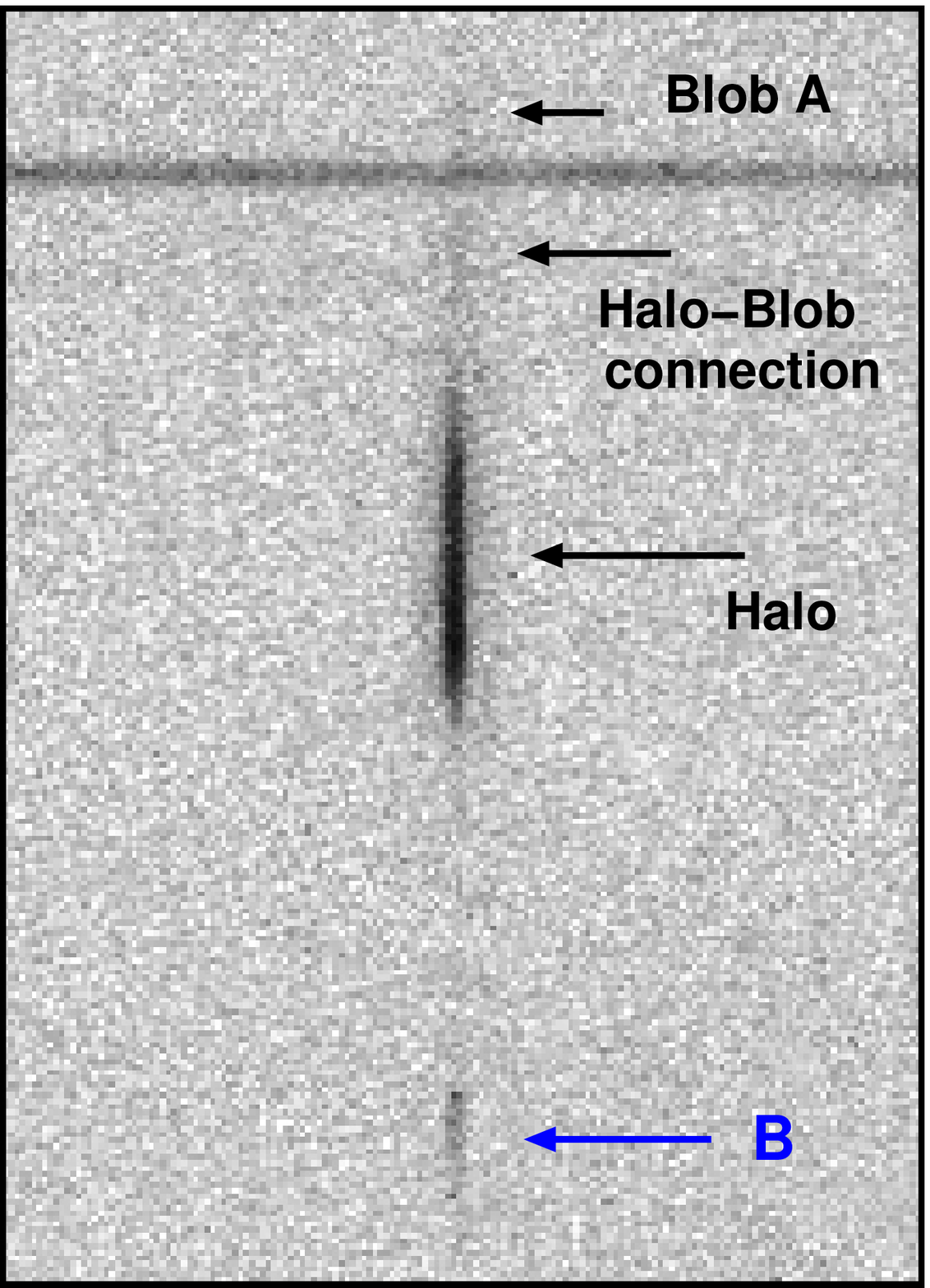}
\vskip .1in
\includegraphics[width=0.935\linewidth]{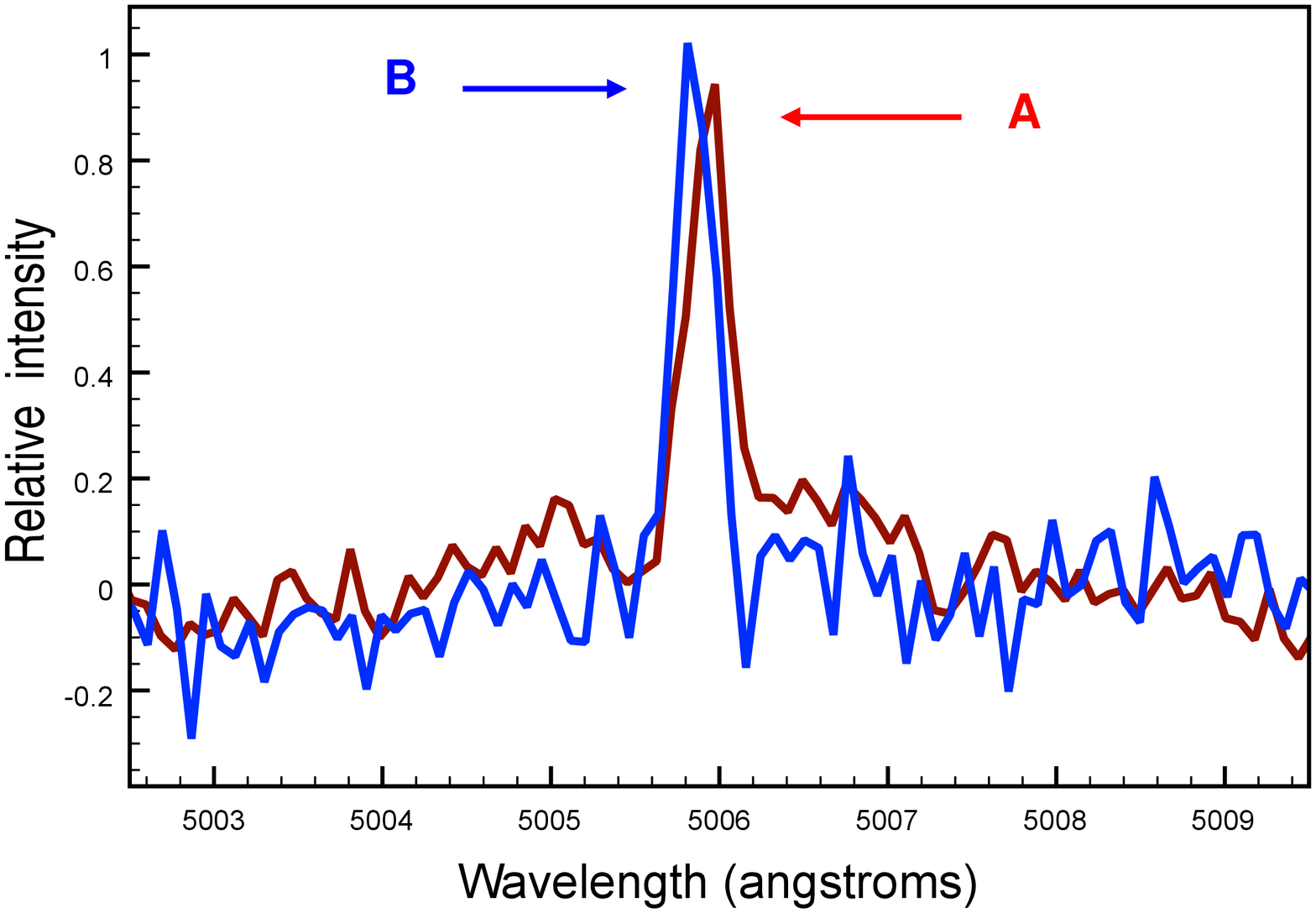}
\caption{
(\emph{top}) Echellograms of slits at PA +64\degr\ across blob A (left) 
and PA +50\degr\ across blob B and halo (right). The blob A is barely visible in the right image because the slit passes slightly below the blob (see Figure~\ref{NOTlarge}.
The spatial location of the blobs, halo and halo-blob connection are labeled.   
(\emph{bottom}) One-dimensional averaged spectra of blobs A and B corresponding at 28$\arcsec$ and centred at the peak emission were extracted from the echellograms in top panels. 
\label{spectra}
}
\end {center}
\end{figure}

\section{Results}
\label{results}

We have in hand kinematical and spatial information of the main structural 
components of NGC 6309: the bipolar lobes and central ring, the halo, and the 
outermost blobs.  
This information can be used to derive the kinematical ages of each 
component in order to investigate their formation history.  
In the next subsections, we consider each component in turn.  

\subsection{The bipolar lobes: spatio-kinematical modelling}
\label{shape}

The central regions of NGC\,6309 were studied in detail by V08 using 
narrow-band images and high-dispersion echelle spectra.  
They concluded that the inner bright structure can be interpreted as an 
expanding torus and two pairs of bipolar lobes oriented along PAs 40\degr\ 
and 72\degr.  
These structures are revealed with unprecedented detail in the 
high-resolution images presented in Figure~\ref{NOTzoom}.  
The central ring, which is the brightest structure, seems to be formed by 
a non-uniform assembly of knots especially prominent in [N~{\sc ii}], that 
can be fairly fitted by an ellipse 20\arcsec$\times$8\arcsec\ in size.  
The conical eastern structure reported by V08 after the use of 
``unsharp-masking'' techniques is plainly evident in our direct 
images.  
It connects the ring-like central structure to the northeast arm 
described in old, low resolution imagery.  
A possible western counterpart is absent.  
The bipolar lobes are composed by small gaseous `flakes' and their tips 
display a characteristic wave-like pattern.  
The two pairs of bipolar lobes, which were discovered by their distinct 
kinematics using long-slit echelle data (V08), are neither obvious 
features, but they can be hinted in this direct image.  

\begin{figure*}
\begin{center}
  \includegraphics[width=\textwidth]{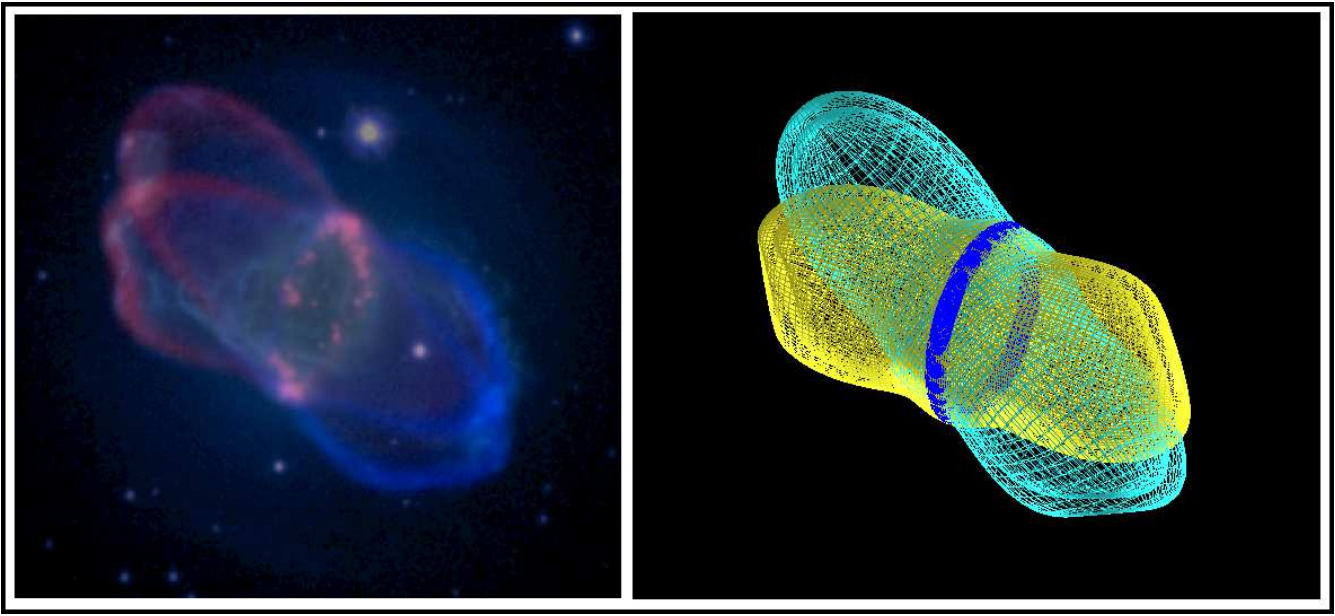}
  \caption{
Structure model of NGC\,6309 produced with {\sc shape}.  
Left: Model superimposed on the composite RGB image.  
Right: The model formed by two bipolar outflows and an expanding ring.  
}
  \label{shapemodel}
\end {center}
\end{figure*}

Using the simple hourglass model of \citet{sol85}, and assuming a distance of 2 kpc, V08 derived kinematical 
ages $\sim$4000 yrs and $\sim$3700 yrs for the bipolar lobes along 
PA 40\degr\ and 72\degr, respectively.  
The inclination angle with respect to the line of sight was estimated 
to be $66\degr$ from the assumption that the axis of symmetry of the lobe at PA 72\degr
 coincides with the main axis of the ring, for a deprojected expansion velocity of the ring of 25 
km\,s$^{-1}$.
Using the software {\sc shape} \citep{ste11}, specifically designed for 
the spatio-kinematical modeling of expanding nebulae, we have fit our 
images and the high-dispersion spectra presented by V08 adopting a 
sophisticated model that includes two irregular bipolar lobes and an 
expanding ring.  
The deviations of the bipolar lobes from Solf \& Ulrich's (1985) prescription 
are modeled using the {\sc bump} modifier in the {\sc shape} software.  
The synthetic image implied by the best-fit model is presented in 
Figure~\ref{shapemodel}. 

The best-fit model parameters listed in Table~\ref{parameters} represent 
small adjustments with respect to those presented by V08.  
In particular, the estimate of the kinematical ages of the bipolar 
lobes remains unchanged.  

\begin{table}
 \caption{Parameters of main structures measured and derived from models. All
 values correspond to deprojected parameters.}

\begin{tabular}{@{}lccc@{}}
\hline
\hline
Parameter                                     & Ring               &  Bipolar outflow  & Bipolar outflow  \\
                                                      & PA 72{\degr}  & PA 40{\degr}       & PA 72 {\degr}  \\
\hline 
Semi-major axis (arcsec)              & 10                  & 35                       &  28 \\
Equatorial velocity (km\,s$^{-1}$) & 25                  & 25                        & 25 \\           
Polar velocity (km\,s$^{-1}$)         & ---                  & 84                        & 72 \\ 
Main axis inclination ({\degr})        & 66                 & 66                        & 66 \\
Kinematical age (yr)                      &  3800            & 3950                    & 3750 \\
\hline
\end{tabular}
\label{parameters}
\end{table}

\subsection{The faint halo}

\begin{figure}
\begin{center}
  \includegraphics[width=0.475\textwidth, angle=0]{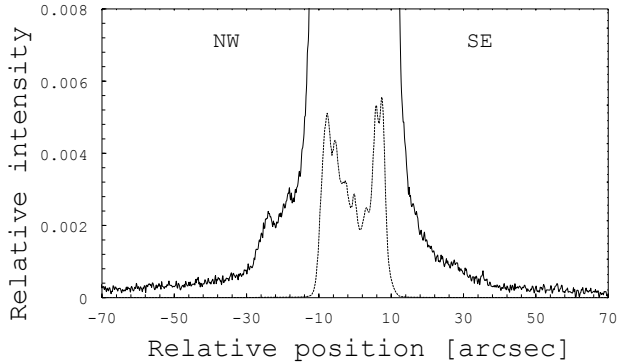}
\caption{[O~{\sc iii}] surface brightness profile of the halo of NGC\,6309
along the Northwest-Southeast direction at PA=--30\degr\ .The profile has been extracted using an aperture of width 1\farcs8 in order
to attain an adequate signal-to-noise ratio.
The halo's profile is normalized to the emission peak.
\label{profile}}
\end {center}
\end{figure}

The new narrow-band images of NGC\,6309 shown in Figs.~\ref{NOTzoom} and
\ref{NOTlarge} clearly unveil a shell surrounding the bipolar lobes.
This shell, which we will recall as the halo of NGC\,6309 following V08, 
has a round
appearance, with a size $\sim$60$\arcsec$.
The shell is only detected in the [O~{\sc iii}] image, while remaining
unseen in the H$\alpha$ and [N~{\sc ii}] images.
The shell shows a distinctive limb-brightened arc towards the Northwest,
whereas it fades progressively along all other directions.
This is illustrated in Figure~\ref{profile}, which shows the halo's
surface brightness profile extracted along the northwest-southeast direction.
The halo surface brightness is $\sim$100 times fainter than the central
region of the nebula.

The presence of this structural component of NGC\,6309 was previously
reported by V08, that described it as a diffuse halo, but could not
provide too much detail.
Actually, it was \citet{chu89} the first who reported the presence of
an extended shell around the bright innermost regions of NGC\,6309
based on echelle data.
They computed an expansion velocity $\sim$5 km~s$^{-1}$ using the line
width from their long-slit echelle spectrum.

It is interesting to revisit this expansion velocity to gain a better
understanding of the evolutionary past of NGC 6309.
The faint halo around the bright bipolar lobes is registered by the
long-slit echelle spectrum covering blob B (Fig.~\ref{spectra}
upper-right panel).
This spectrum shows an unresolved  [O~{\sc iii}] line.
We have computed the line width in space bins of 5 pixels
($\sim$3\farcs5) and found that the FWHM is rather constant,
21.0$\pm$1.5 km\,s$^{-1}$.
After subtract the instrumental and thermal width, as described by \citet{bry92}, it results in a FWHM of 17.2$\pm$1.5 km~s$^{-1}$.
The variance of the line width can be used to derive an upper limit for
the expansion velocity \citep{gue98a} as the line width can be expected
to increase with the path length of the shell along the line of sight.
The shell appearance and its surface brightness imply it is a filled
shell.
Assuming constant emissivity, we have estimated an upper limit for the
expansion velocity along the line of sight of 5.75 km\,s$^{-1}$.
After correcting for projection effects, and assuming a distance of 2 kpc this implies an upper limit to the expansion velocity of $<$6.1 km\,s$^{-1}$ and a lower limit to the expansion age, for a shell radius of 30$\arcsec$, of $>$46,000 yr, comparable to the [O~{\sc iii}] halo expansion velocity of NGC\,6543 \citep{bry92}.

A relevant aspect is that the halo's centre, as determined by an [O~{\sc iii}] 
brightness contour at low intensity, does not line up with the central star.  
This offset suggests that the star presents a proper motion 
along the northeast direction.  
Moreover, the bipolar structure protrudes into the NE edge of the halo, 
which provides another element to support this hypothesis.  
The estimation of the direction of the movement of the star from the 
relative position of the halo centre differs from the northwest direction reported by \citet{ker08}, also, the external halo show a kind of bright non diffuse rim in the NW direction (see Fig.~\ref{NOTlarge}).  
We attribute this inconsistency to the difficulties to determine the 
position of the central star in a region with bright and irregular 
diffuse emission.  

\begin{figure*}
\begin{center}
  \includegraphics[width=0.35\textwidth, angle=270]{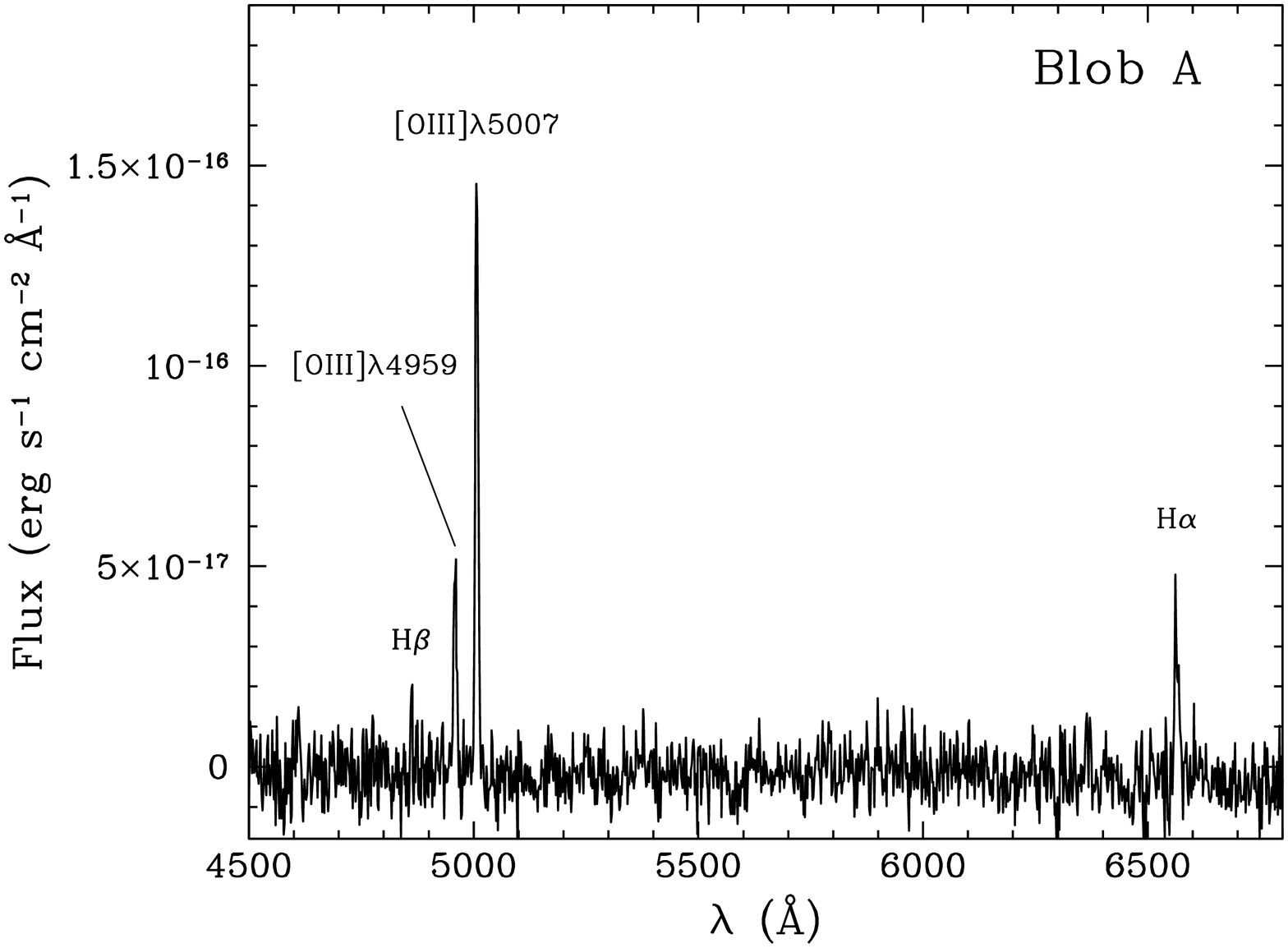}
  \includegraphics[width=0.35\textwidth, angle=270]{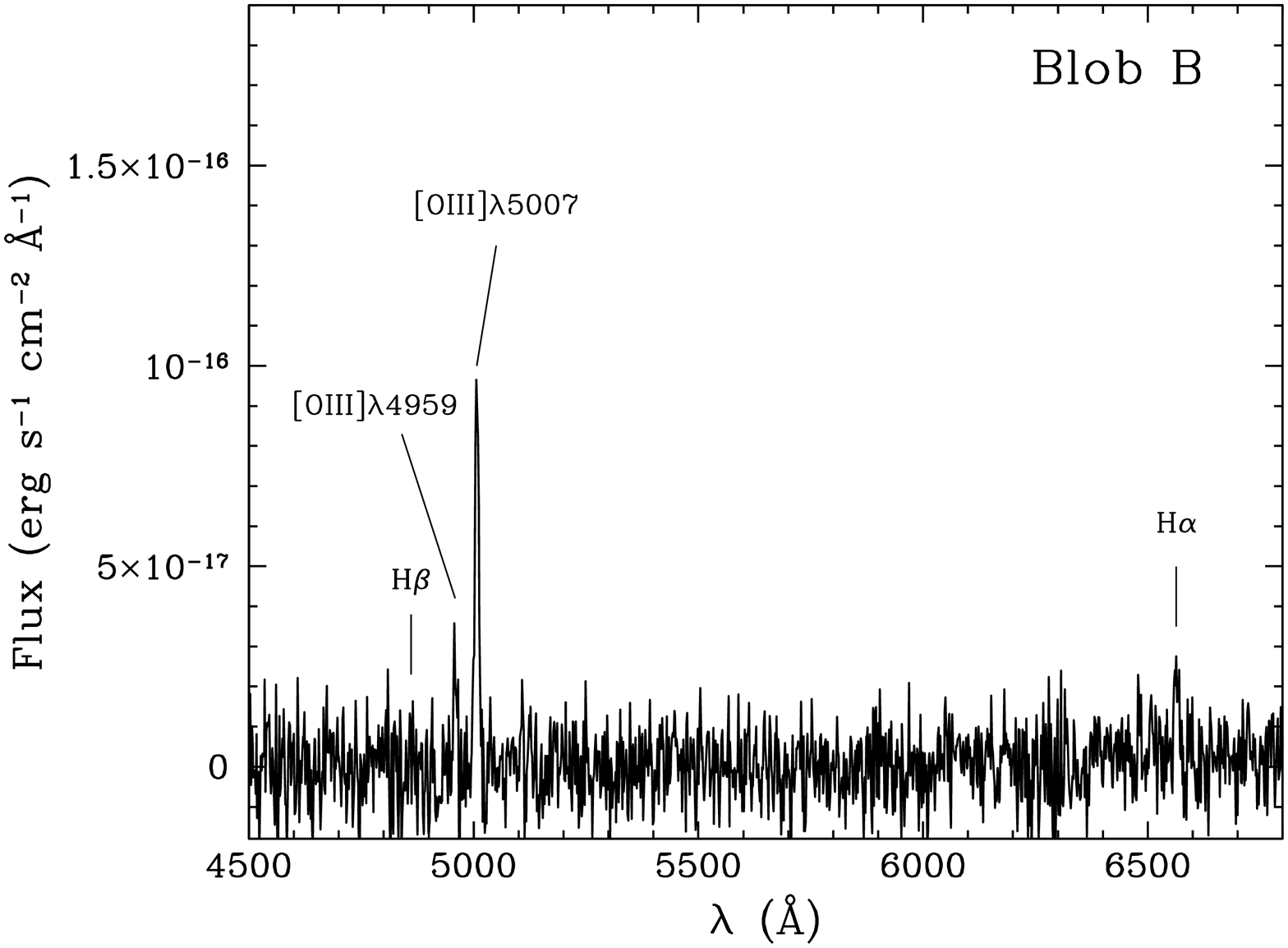}
\caption{One-dimensional spectra extracted from the blobs A and B. 
Aperture regions of 2\farcs6$\times$14\farcs2 and 3\farcs9$\times$17\farcs7 
centred on blob A and B have been used, respectively.
\label{spectrab}}
\end {center}
\end{figure*}

\subsection{The blobs}

The large field of view image of NGC\,6309 presented in Figure~\ref{NOTlarge} 
reveals two faint blobs: the eastern blob ``A'' is located at 55\arcsec\ from 
the CSPN along PA $\approx$70\degr, whereas the western blob ``B'' is located 
at 78\arcsec\ along PA $\approx$210\degr.  
Therefore, they are not located symmetrically with respect to the nebular 
centre. They have a size of 20\arcsec$\times$10\arcsec\ and their appearance 
is diffuse, without any distinctive feature.  

The blobs emit predominantly in [O {\sc iii}], with blob A being brighter 
than blob B. 
The averaged surface brightness is 
1.8$\times$10$^{-13}$ erg~cm$^{-2}$~s$^{-1}$~arcsec$^{-2}$ for blob A and 
8.3$\times$10$^{-14}$ erg~cm$^{-2}$~s$^{-1}$~arcsec$^{-2}$ for blob B. 
Spectra of both blobs are shown in Figure~\ref{spectrab}. 
The  [O\,{\sc iii}] $\lambda\lambda$4959,5007 emission lines can clearly 
be seen in both blobs. 
In the brightest blob A, faint H$\alpha$ and H$\beta$ emission was also 
detected, while in blob B H$\alpha$ was detected, but H$\beta$ 
appears marginal. 
In general blob B shows fainter emission at all the lines.      

\begin{figure}
\begin{center}
  \includegraphics[width=0.48\textwidth, angle=0]{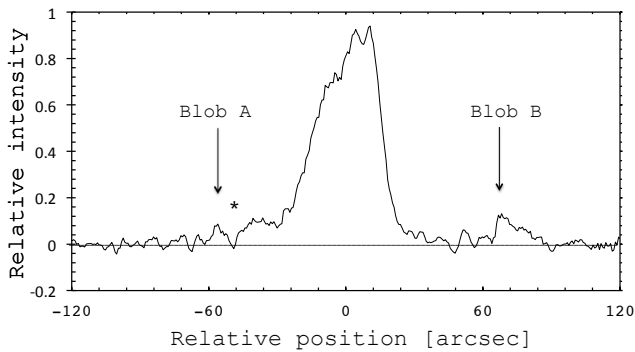}
\caption{Intensity profile extracted from the echellogram corresponding to the slit at PA=+50\degr (Figure~\ref{spectra}, upper-right panel). 
The link between the halo and blob A is clearly seen. The position of an over-extracted star is shown.  
\label{profilehb}}
\end {center}
\end{figure}

The low-dispersion spectra (Figure~\ref{spectrab}) imply [O~{\sc iii}] to 
H$\beta$  ratios similar to those reported by V08 for the central regions.  
This seems to suggest that the blobs are excited by the UV flux of the 
central star. On the other hand, the H$\alpha$ to H$\beta$ ratio is smaller than in the 
central nebula, indicating lower extinction.  
Indeed, a logarithmic extinction coefficient c$_{\mathrm{H}\beta}$ of 0.25 
is derived for blob A, whereas for the central region of NGC\,6309 it is 
estimated to be 0.90 \citep{vaz08}.  
 
The echelle spectra provide valuable information on the kinematics of 
these blobs.  
The one-dimensional averaged spectra extracted for each blob reveal a 
small although significative difference in their radial velocities 
(Figure~\ref{spectra}).  
The peak emission of blob A is redshifted, with a systemic velocity\footnote{
The radial velocity is referred to the systemic velocity, as derived 
by V08, $V_\mathrm{LSR}=-32\pm2$ km\,s$^{-1}$.  
} 
$V_\mathrm{A}$=+3 km\,s$^{-1}$, whereas the emission peak of blob B is blueshifted, $V_\mathrm{B}$=-4 km\,s$^{-1}$.   
The two-dimensional echellograms show the blobs to be almost inert, with 
internal velocity variations within the line-width, i.e., without significant 
kinematical structure.  

A close inspection of the echellogram along blob B 
(Fig.~~\ref{spectra}-top-right) reveals weak emission connecting blob A 
to the halo of NGC\,6309.  
This connection is clearly shown in the intensity profile of the 
[O~{\sc iii}] extracted from this echelle spectrum (Figure~\ref{profilehb}).  
Blob A is thus not isolated, but it is connected to the halo of NGC\,6309.  
This physical connection and the similarity between the radial velocity of 
the blobs and that of NGC\,6309, and their excitation degree strongly 
support the association of the blobs with NGC\,6309.  

A deep [O {\sc iii}] image (second observation, see Section 2.1) reveals weak 
flocculent structures in the southern part of the halo.  
These features, not shown here, are not completely undisputed, as we 
recognize they may be caused by light dispersed from the bright inner 
nebular by different optical elements of the telescope and camera 
\citep[see][for a thorough discussion on the difficulties of detecting faint nebular features around bright PNe]{cor03}. 
 
\section{Discussion}  
\label{discusion}

The quality of the narrow-band images of NGC\,6309 presented in this 
work has allowed us to obtain a precise description of morphological 
features that were only hinted in previous studies (the eastern conical 
structure, the round halo) or simply undetected (the outer faint blobs).  
The spectroscopic information gathered here can be used to shed 
light into the origin of these structures and to investigate the 
formation history of this PN.  

\subsection{The origin of the faint outer blobs}

The nature of the faint outer blobs A and B is really intriguing.  
The spatial, spectroscopic, and kinematical information provided in 
Section 3.2 can put to the test different scenarios to assess 
their origin.  

It is certainly appealing to identify blobs A and B with high-velocity 
polar ejections.  
High-velocity knots located at several or many radii of the main nebular 
shell have been identified in a number of PNe, e.g., Fleming\,1 and 
MyCn\,18 \citep[][]{lop93,bry97}. 
In some cases, there is evidence of the interaction of these outflows 
with the nebular shell \citep[e.g., NGC\,6778,][]{gue12}.
Blobs A and B may have interacted with the bipolar lobes of NGC\,6309: 
blob A is aligned with the eastern conical structure and the CSPN, 
whereas blob B and the CSPN are aligned with an opening in the 
southwestern region of the bipolar lobe.  
This would imply that the blobs, having played a role in the formation 
of the bipolar lobes \citep{sah98}, are coeval to those or 
followed immediately their formation.  
Since the bipolar lobes have expansion velocities $\sim$80 km~s$^{-1}$ 
and angular extents 2--3 times smaller than the radial distances of 
blobs A and B to the CSPN, we can infer expansion velocities of these 
blobs in excess of 160--240 km~s$^{-1}$.  
As their systemic velocity is only 3--4 km~s$^{-1}$, the inclination 
angle with respect to the plane of the sky must be $\approx$ 1 \degr.  
This seems very unlikely, especially because blobs A and B are oriented 
along different directions.  
Furthermore, the diffuse appearance of the blobs is in sharp contrast 
with the bow-shock \citep[e.g., IC\,4694,][]{gue08} or compact, 
knotty morphologies \citep[e.g., KjPn\,8,][]{lop97} exhibited by fast outflows 
of PNe.  
We therefore disregard the idea that blobs A and B are fast collimated 
outflows.  

Alternatively, blobs A and B can be interpreted as material of the ISM 
which is illuminated and ionized by the CSPN of NGC\,6309.  
Their excitation degree is high, with [O {\sc iii}] to H$\beta$ ratios 
similar to those of the inner nebula, thus supporting this hypothesis.  
If so, the alignment of these blobs with particular nebular features 
suggests that they correspond to openings in the nebular shell that 
allow a greater UV flux to escape.  
The lower extinction of blobs A and B, as compared to that of NGC\,6309, 
suggests that the main nebular shell suffers from internal obscuration.  

It is interesting to note that the radial velocity of these blobs and 
that of NGC\,6309 main nebular shell are very similar.  
Whereas this does not rule out the possibility that blobs A and B belong to the ISM, 
it hints at a real physical connection between the blobs and the PN.  
This link is strengthened by the emission connecting blob A with 
the halo of NGC\,6309.  
The two blobs can be part of a large and faint structure around NGC\,6309 
ejected at low speeds in the early evolution of the PN \citep{olo00}.  
Assuming a distance of 2 kpc, and an expansion velocity of 5 km~s$^{-1}$, consistent with their 
radial velocity, a kinematical age of 150,000 yrs can be derived.  
This would place this ejection in the late thermal pulse phase of the 
AGB \citep[][]{vas93,sta95}.  
The blobs would be the brightest regions of this structure or sections 
that are illuminated preferentially by the CSPN through holes in the 
nebular shell.  
Examples of incomplete haloes \citep[e.g., IC\,4593,][]{cor97}  or 
inhomogeneous haloes \citep[e.g., NGC\,6543,][]{mid89} abound in the literature.

 \subsection{NGC\,6309: a case study of the change in mass-loss geometry 
in a PN}

The formation history of NGC\,6309 is certainly exceptional.  
Some 150,000 yrs ago, its central star seemed to undergo a mass-loss 
event which resulted in a large, slowly expanding shell-like structure.  
Only the brightest regions of this structure or those regions that 
are preferentially illuminated by the CSPN have been detected so far 
as blobs A and B.  
The long time-scale and slow expansion velocity can be used to identify 
this ejection with one of the last thermal pulses experienced by 
the progenitor star in the late AGB phase.  

More recently, $\sim$45,000 yrs ago, the star experienced another 
episode of heavy mass-loss that resulted in the formation of its 
halo.  
This mass ejection had a clear spherical symmetry which is only distorted 
by the possible interaction of the nebula with the ISM, producing the  
arc-like feature along the direction of the motion of the CSPN.  
The halo may correspond to the very last thermal pulse in the late AGB phase.  

The transformation of the AGB star into a PN occurred $\sim$4,000 yrs ago.  
The star suffered a heavy mass-loss episode that ejected its envelope and 
produced the bipolar lobes.  
At some moment between the ejection of the nebular halo and that of 
the bright inner nebula, the geometry of the mass-loss experienced 
a dramatic change, from spherically symmetric to axially symmetric, 
with rapid changes in the symmetry axis that produced two pairs of 
bipolar lobes along different directions.  
The change in mass-loss geometry experienced by the progenitor of 
NGC\,6309 is consistent with the mechanism proposed by \citet{sah98} 
to explain the onset of asymmetry: fast ($\sim$100 km~s$^{-1}$) 
collimated outflows ejected in the late AGB phase carve through the 
AGB envelope to produce polar or multi-polar impressions in the previously 
spherical envelope.  
The onset of the fast stellar wind will bore through these openings 
in the nebular envelope to shape axially symmetric PNe.  

This scenario is in agreement with the presence of round haloes detected 
in bipolar proto-PNe such as IRAS\,13557$-$6442, IRAS\,17253$-$2831, 
IRAS\,17440$-$3310, IRAS\,19306$+$1407 \citep{sah07a}, and in the 
quadrupolar proto-PN IRAS\,19475$+$3119 \citep{sah07b}.  
Whereas the inner bipolar nebula and outer halo configuration seems common 
in early stages of PN formation \citep{su04}, there is little evidence in 
evolved bipolar PNe.  
Besides the H$_2$ halo detected around NGC\,2440 \citep{ram09}, there are no other bipolar PN surrounded by a round halo
\citep[e.g.,][]{chu87,cor93}.  
Round rings and arcs have been detected around the central regions of 
Hb\,5, NGC\,6881 and NGC\,7026 \citep{cor04}, but their sizes are smaller than the bipolar nebulae and they tend to have significantly
weaker surface brightness than that of the nebular region.
In this sense, the presence of a distinct spherical halo around a 
quadrupolar PN makes of NGC\,6309 a rare case in the bestiary of 
axially-symmetric PNe.

\section{Concluding remarks}

We have presented new observations of NGC\,6309 with deep, high-resolution 
images that reveal its internal structure in great detail, besides confirming the presence of a spherical halo. 
Likewise, we show unprecedented structures outside the main nebula, 
which we have termed blobs.  
We have carried out high- and low-dispersion spectroscopy 
to better understand the nature and kinematics of these 
morphological features.  

NGC\,6309 has a remarkable morphology and allows us to clearly see the 
record of different mass-loss episodes; the kinematics of the blobs reveals 
that these components probe an early mass ejection which took place during 
the thermal pulse phase, $\approx$150,000 yrs ago.  
The surrounding halo, seen clearly for the first time in our images, is the 
remnant of symmetrical mass-loss in the final stages of the AGB phase.  
The picture is completed with two pairs of bipolar lobes created during 
a short period 4,000 yrs ago.
We emphasize the change in mass-loss geometry from the final stages 
of the AGB, when the round halo was ejected, to the early PN phase, 
when multiple bipolar lobes were ejected along different directions.  

In order to investigate the origin of the late changes in mass-loss in 
NGC\,6309, an exhaustive study of its CSPN is required to search for 
variability or chemical abundances anomalies associated to its evolution 
through a common envelope phase.  
A study of dust polarization emission can also shed light in the search 
for magnetic fields.  
The study of the late evolution and formation of bipolar PNe will also 
benefit from the understanding of the low occurrence of round outer 
structures among these sources.

\section*{Acknowledgments}

We are grateful to the staff of OAN-SPM, specially to Gustavo Melgoza-Kennedy, 
telescope operator, for his assistantship during observations. 
This paper has been supported by grant PAPIIT-DGAPA-UNAM IN107914.
GR acknowledges CONACYT for his graduate scholarship and 
the IAA for its hospitality during his stay.
GR-L acknowledges support from CONACYT (grant 177864), CGCI, PROMEP and SEP (Mexico).
MAG also acknowledges support from grant AYA 2011-29754-C03-02.

\label{lastpage}

\end{document}